\documentclass[%
 reprint,
 amsmath,amssymb,
 aps, prd,preprintnumbers, superscriptaddress, nofootinbib
]{revtex4-2}

\usepackage{graphicx}	
\usepackage{amsmath}	
\usepackage{amssymb}	

\usepackage[english]{babel}
\usepackage{graphicx}
\usepackage{latexsym}
\usepackage{floatflt}
\usepackage{float}
\usepackage{units}
\usepackage{hyperref}
\usepackage{amsmath}
\usepackage{gensymb}
\usepackage{mathdots}
\usepackage{color}
\usepackage{xcolor}
\usepackage{mathtools}
\usepackage{lipsum}
\usepackage{slashed}

\usepackage{dcolumn}
\usepackage{bm}

\newcommand{\id}{{\rm d}}

\def\fd{f_{\rm D}}

\def\mpd{m_{\rm p_{D}}}
\def\med{m_{\rm e_{D}}}
\def\alphad{\alpha_{\rm D}}
\def\qd{q_{\rm D}}
\def\mHd{m_{H_{\rm D}}}
\def\mud{\mu_{\rm D}}
\def\xid{\xi_{\rm D}}
\def\zres{z_{\mathrm{res}}}
\def\mgd{m_{\gamma_{\rm D}}^{2}}
\def\mg{m_{\gamma}^{2}}

\begin{document}

\preprint{MIT-CTP/6003}
\preprint{YITP-SB-2026-04}

\title{CMB Spectral Distortions from Resonant Conversions in Atomic Dark Sectors}
\author{Duncan K. Adams}
\email{duncan.adams@stonybrook.edu}
\affiliation{C.N. Yang Institute for Theoretical Physics, Stony Brook University, Stony Brook, NY 11794-3800, USA}

\author{Jared Barron}
\email{jared.barron@stonybrook.edu}
\affiliation{C.N. Yang Institute for Theoretical Physics, Stony Brook University, Stony Brook, NY 11794-3800, USA}

\author{Bryce Cyr}
\email{brycecyr@mit.edu}
\affiliation{Center for Theoretical Physics - A Leinweber Institute, Massachusetts Institute of Technology, Cambridge, MA 02139, USA}

\author{Xiuyuan Zhang}
\email{xiuyuan@mit.edu}
\affiliation{Kavli Institute for Astrophysics and Space Research, Massachusetts Institute of Technology, Cambridge, MA 02139, USA}

\begin{abstract}
Dark sectors consisting of atomic constituents (electrons, protons, and photons) offer a well-motivated extension to the Standard Model while providing multiple avenues for phenomenological study. In this work, we explore the impact of conversions between the dark and Standard Model photons in the primordial CMB spectral distortion epoch ($10^3 \lesssim z \lesssim 10^6$). These conversions are resonantly enhanced when the induced thermal masses of both photonic species are equal, thus leading to the possibility that sizeable distortions can be produced. To this end, we solve the Boltzmann equation at early times to determine the (irreducible) freeze-in or freeze-out abundance of dark photons. This procedure also allows us to update the limits on generic milli-charged dark sectors using the ACT DR6 bound on the number of effective radiative degrees of freedom ($N_{\rm eff}$). By then modeling the evolution of the thermal masses in both sectors, we compute the primordial CMB distortion using the Landau-Zener formalism. We find that when the dark electron and proton are roughly similar in mass (the positronium limit), current spectral distortion data from the COBE/FIRAS instrument is able to rule out novel regions of parameter space. We also forecast bounds from the proposed FOSSIL satellite, finding that spectral distortions can also be used to probe the ultra-low dark electric charge regions of parameter space, which are difficult to investigate by other means.
\end{abstract}

\maketitle

\section{\label{sec:level1}Introduction}
Additional gauge symmetries are ubiquitous in extensions to the Standard Model of particle physics. The most simple models posit the existence of a single new $U(1)_{\rm D}$ symmetry, giving rise to a gauge boson often referred to as the dark photon \cite{Holdom1985}. In these minimal scenarios, the dark photon is typically coupled to the Standard Model through a kinetic mixing with the visible photon, whose interaction strength is governed by the free parameter $\epsilon$. For sufficiently feeble interactions, this new gauge boson is capable of serving as the entirety of the dark matter for a wide range of dark photon masses (see \cite{Fabbrichesi2020, Caputo2021} for recent reviews\footnote{As well as \url{https://cajohare.github.io/AxionLimits/docs/dp.html} for up-to-date limits.}). 

If the dark photon comprises the entirety of dark matter, it must either be sufficiently massive such that it is non-relativistic during structure formation (cold), or it must be produced non-thermally, in which case one is free to consider arbitrarily low masses. In the massless limit however, there is no longer any notion of a \textit{cold} dark photon, and they are often considered as part of a more complex theory. This becomes necessary for observational signatures, as in a pure $U(1)_{\rm D}$ setup the mixing between the dark and visible sectors can be rotated away by a field redefinition. When this is the case, the dark photon decouples\footnote{Unless one considers operators with mass dimension $>4$.} from the Standard Model, and the dominant observational effects come from an effective milli-charge between the Standard Model photon and the dark current ($j_{\rm \mu, D}$) of the form $\mathcal{L}_{\rm int} \sim  \epsilon j_{\mu, \rm D} A^{\mu}$. Thus, for massless dark photons, the presence of fermions charged under $U(1)_{\rm D}$ in the model provides a reasonable interaction channel with the visible sector and the potential to thermally produce dark radiation.

Massless dark photons are extra radiative degrees of freedom ($\Delta N_{\rm eff}$), and can therefore be constrained through their indirect impacts on light element abundances produced during big bang nucleosynthesis (BBN) \cite{Cyburt2015, Fields2019, Schoneberg2024}, as well as the structure of the acoustic peaks in the cosmic microwave background (CMB) power spectrum \cite{Planck2018, Yeh2022}. Combining astrophysical observations of the light elements with recent CMB observations from the Atacama Cosmology Telescope (ACT DR6) and the Planck satellite, one finds $N_{\rm eff} = 2.89 \pm 0.11$ ($1\sigma$) \cite{ACT2025}. Limits on the massless dark photons in these milli-charged models can thus be obtained by requiring that their abundance does not violate the bound on $\Delta N_{\rm eff}$~\cite{Vogel:2013raa, Dolgov2013,Foot:2014uba,Creque-Sarbinowski2019, Adshead:2022ovo}. 

Interestingly, the inclusion of a background of dark fermions can also generate an effective plasma (thermal) mass for the dark photons. The endowment of a dark photon mass is notable, as it opens up additional phenomenology that can allow for new observational channels into the dark sector. One such channel is the possibility for resonant conversions between dark and Standard Model photons, which can occur when the plasma masses of both the visible and dark photons are equal. If these conversions take place in the primordial distortion epoch ($10^3 \lesssim z \lesssim 10^6$), characteristic $\mu$- and $y$- type patterns can be searched for in CMB spectral data.

At $z\lesssim 10^6$, the lowest order photon number changing interactions (double Compton scattering and Bremsstrahlung) freeze-out in the standard cosmological model. After this point, perturbations of the photon frequency spectrum (through e.g. the injection or extraction of energy and entropy) away from a pure blackbody can no longer be brought back into full thermal equilibrium, thus marking the onset of the CMB spectral distortion epoch. In the case of a massive dark photon, resonant conversions both into and out of the dark sector can generate significant spectral distortion signatures, providing leading constraints on vanilla (e.g. no dark fermions) dark photon models for masses in the range of $10^{-14} \, {\rm eV} \lesssim m_{\gamma_{\rm D}} \lesssim 10^{-4} \, {\rm eV}$ \cite{Mirizzi2009, Arias2012, McDermott2019, Caputo2020, Chluba2024, Arsenadze2024}. 

For the case of massless dark photons, Berlin \textit{et al.} \cite{Berlin:2022hmt} found that constraints could still be set provided that fermions charged under the dark $U_{\rm D}(1)$ are present. As mentioned above, the presence of dark fermions gives rise to a plasma mass for the dark photon, restoring the ability for resonant conversions to take place at the critical redshift(s) when $m_{\gamma} \simeq m_{\gamma_{\rm D}}$. At low frequencies and for non-relativistic fermions, the thermal mass (plasma frequency) in the dark (visible) sector scales with $m^2_{\gamma_{\rm D}} \sim n_{\rm e_D}/m_{\rm e_{D}}$ where $n_{\rm e_D}$ is the number density of free dark fermions (electrons). Due to fact that the redshift dependence on the thermal mass in both sectors is identical, the authors of \cite{Berlin:2022hmt} found that the resonance condition could only be satisfied at times when the free-electron fraction went through a dramatic change, namely, at recombination and reionization. It is, however, possible to consider a scenario in which the matching occurs in the primordial spectral distortion epoch ($10^3 \lesssim z \lesssim 10^6$), allowing us to make full use of the constraining power provided by the COBE/FIRAS experiment \cite{Fixsen1996, Fixsen2003, Bianchini2022, Sabyr2025}. To do so we extend the analysis of Berlin \textit{et al.} and consider an atomic dark sector in which the dark plasma undergoes its own recombination at these early times. 

The main results of this work are comprehensive constraints from the CMB on atomic dark sectors with a kinetically mixed dark photon stemming from a combination of $\Delta N_{\rm eff}$ and spectral distortions. In addition, we determine regions of parameter space that are clear targets for upcoming spectral distortion experiments such as TMS \cite{TMS}, BISOU \cite{Maffei2021}, COSMO \cite{Masi2021}, and the currently proposed space-based mission FOSSIL \cite{2024EPJWC.29300012C}. As a secondary result of this work, we provide updated constraints on generic milli-charged dark sectors from the most recent ACT bounds on $\Delta N_{\rm eff}$, yielding modest improvements over previous results \cite{Vogel:2013raa, Adshead:2022ovo}.

The remainder of the paper is organized as follows: In Section \ref{sec:level2} we provide an overview of the construction of an atomic dark sector, including a description of the production mechanisms considered, and the subsequent evolution of the phase space distributions through the relevant Boltzmann equation, allowing us to fully characterize the contribution to $\Delta N_{\rm eff}$. We provide an expression for the thermal masses of the hidden and visible photon (Sec.~\ref{sec:level2.2}), and discuss details of the dark recombination process (Sec.~\ref{sec:level2.3}) following a Saha approach. Following this, Section \ref{sec:level3} is dedicated to the calculation of the spectral distortion signature, comparing and contrasting the bounds achievable using $\Delta N_{\rm eff}$ alone. We discuss the results and conclude in Sec.~\ref{sec:level4}. Details of the Boltzmann setup can be found in Appendix~\ref{sec:appA}. Throughout, we use natural units in which $c = \hbar = k_{\rm b} = 1$, and our Fourier transform follows the convention $A(\textbf{x},t) = \int \id^3 \textbf{k}\, \id  \omega \, \tilde{A}(\textbf{k},\omega) {\rm e}^{i(\textbf{k}\cdot \textbf{x} - \omega t)}/(2\pi)^{3/2} $.

\section{\label{sec:level2}Atomic Dark Sectors}

Efforts to restore symmetry or reduce tuning in the Standard Model have fostered theoretical interest in models with complex hidden sectors. Since the 1980s, the mirror-world scenario which was originally envisioned by Yang and Lee to restore parity symmetry \cite{Lee:1956qn} has been considered for its cosmological implications, including as a model of dark matter \cite{Blinnikov:1982eh,Blinnikov:1983gh,Kolb:1985bf,Goldberg:1986nk,Khlopov:1989fj,Hodges:1993yb,Berezhiani:1995am,Foot:1999hm,Foot:2000vy,Mohapatra:2000qx,Mohapatra:2001sx,Foot:2002iy,Berezhiani:2003wj,Berezhiani:2003xm,Foot:2003eq,Foot:2003jt,Foot:2004pa,Berezhiani:2005ek}. Efforts to ameliorate the hierarchy problem in the absence of electroweak-scale supersymmetry led to the development of models of Neutral Naturalness, beginning with the Mirror Twin Higgs model \cite{Chacko:2005pe,Chacko:2005vw,Craig:2015pha,Craig:2016lyx}.
Both of these scenarios introduce copies of Standard Model particles and gauge forces in a secluded hidden sector. The hidden sector counterpart of hydrogen could then be stable and form some or all of the dark matter. 

A pared-down version of this scenario containing only the hidden sector partners of the proton, electron, and photon was introduced in \cite{Kaplan:2009de}, named atomic dark matter~ \cite{Kaplan:2009de,Kaplan:2011yj,Cline:2012is,Cyr-Racine:2012tfp}. The model consists of an unbroken $U(1)_{\rm D}$ gauge symmetry, along with two Dirac fermions which have equal and opposite charges under the $U(1)_{\rm D}$, which we take to be 1 and -1. The dark photon is massless in this scenario, while the fermions possess elementary charge $e_{\rm D}$, yielding a dark fine structure constant $\alphad = e_{\rm D}^2/4\pi$. In analogy with the Standard Model, the two dark fermions are called the dark proton and dark electron, with the heavier of the two identified as the dark proton. In contrast to the usual proton, we treat both of these fermions as fundamental particles. In particular we do not assume the existence of a dark QCD, so there is no `nuclear physics' in the dark sector. Through their interaction with the dark photon, the dark proton and dark electron can form neutral bound states which we call dark hydrogen. The Lagrangian of this dark sector is as follows: 
\begin{equation}
\label{eq:Lagrangian}
    \mathcal{L}_{\rm aDM} = -\frac{1}{4}A_{\mu\nu}A^{\mu\nu} + \bar{\tilde{p}}_{\rm D}(i\slashed{D}-\mpd)\tilde{p}_{\rm D} + \bar{\tilde{e}}_{\rm D}(i\slashed{D}-\med)\tilde{e}_{\rm D},
\end{equation}
where $A_{\mu\nu}$ is the field strength tensor for the dark photon, and $\tilde{p}_{\rm D}$ and $\tilde{e}_{\rm D}$ are the dark proton and electron fields. It is assumed that some dark baryogenesis mechanism is responsible for an asymmetric relic abundance of both fermions, such that at late times there are equal numbers of dark protons and dark electrons. The atomic dark matter is assumed to account for a fraction $\fd \equiv \Omega_{\rm aDM}/\Omega_{\rm DM}$ of the dark matter, with the rest being cold and collisionless. The thermal distribution of dark photons is assumed to have a temperature today of $T_{\rm D}$, with a ratio today of $\xid \equiv T_{\rm D}/T_{\rm SM}$ relative to the CMB. The minimal atomic dark matter model thus has five parameters: $\fd$, $\xid$, $\mpd$, $\med$, and $\alphad$. 

With the addition of a kinetic mixing term $-\frac{\epsilon}{2}F_{\mu\nu}A^{\mu\nu}$ between the dark and SM photons, this parameter space is extended by the mixing $\epsilon$. After rotating away the kinetic mixing term, the dark fermions interact with the SM photon with milli-charge $\qd\equiv \epsilon e_{\rm D}/e$. The addition of this coupling between the visible and dark sectors opens channels for the transfer of energy and entropy between the two, such that even an initially cold, unpopulated bath of dark photons is unavoidably heated to some minimum temperature. Depending on the milli-charge $q_{\rm D}$ and masses of the dark fermions, the neutrino-to-photon temperature ratio can also be altered from its usual SM value, breaking the one-to-one relationship between $\xid$ and $\Delta N_{\mathrm{eff}}$. In other words, the dark photon temperature ratio $\xid$, or equivalently the total contribution to relativistic energy density $\Delta N_{\mathrm{eff}}$, can no longer be set arbitrarily small in the theory.

The values of both $\xid$ and $\Delta N_{\mathrm{eff}}$ are critical for determining whether a given aDM model is consistent with observations or not. The temperature ratio $\xid$ strongly impacts the redshift at which dark hydrogen recombines and the atomic dark matter decouples from the dark photons. The later this decoupling occurs, the more the CMB and matter power spectra are modified relative to a fully CDM dark sector~\cite{Cyr-Racine:2012tfp,Cyr-Racine:2013fsa,Foot:2016wvj,Bansal:2022qbi,Barron:2025dys,Barron:2026nks}. Models consistent with these observations generally undergo dark recombination much earlier than the visible sector. Also, as we will see in Section~\ref{sec:level2.3}, the crossing of the dark photon and photon plasma masses---and the resonant conversion between the sectors---occurs at the time of dark hydrogen recombination. Any resulting spectral distortion is therefore dependent on $\xid$. Separately, BBN and the CMB power spectrum provide constraints through $\Delta N_{\mathrm{eff}}$. 

In order to make realistic predictions of the spectral distortion signal and correctly identify constrained parameter space, we must compute $\xid$ and $\Delta N_{\mathrm{eff}}$ as a function of the aDM model parameters. We assume that the dark photons start with negligibly low temperature. In cases where the dark photons never achieve thermal equilibrium with the SM photons, our results should be treated as a lower bound on the final dark photon temperature, because higher initial dark photon temperatures are also possible. However, in cases where the dark photons are brought into thermal equilibrium with the SM photons at some time, our predictions for late-time values of $\xid$ and $\Delta N_{\mathrm{eff}}$ are insensitive to initial conditions. We will find that in the parameter space newly constrained by spectral distortions, thermal equilibrium between the sectors is achieved, making our results general with respect to the abundance of dark photons at very early times. 

\begin{figure*}[t]
    \centering
    \includegraphics[width=1.0\textwidth]{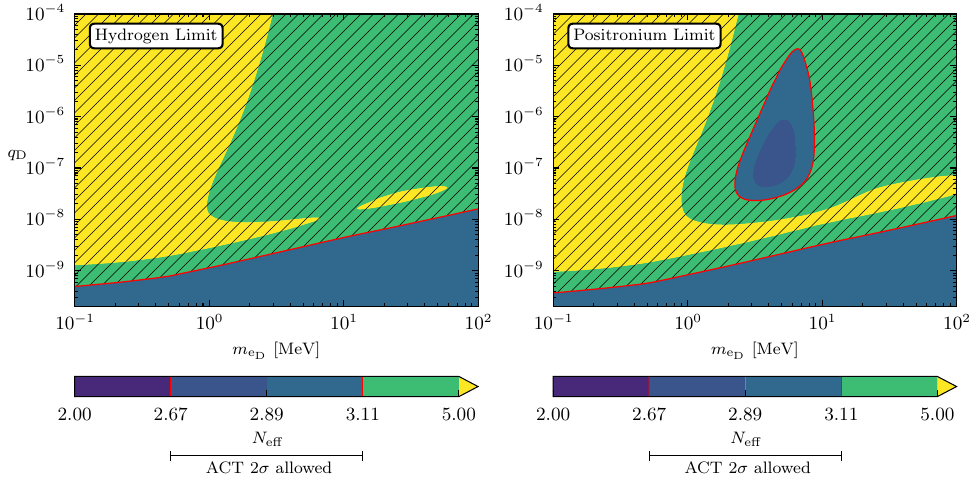}
    \caption{$N_{\rm eff}$ contours in the $m_{\rm e_D}-q_{\rm D}$ parameter space for the hydrogen limit (left) and positronium limit (right). In each panel, the hatched areas indicate sections of the parameter space that are constrained from the combined Planck+ACT+LSS+BBN 95\% CL \cite{ACT2025} ($N_{\rm eff} > 3.11$), with the red contour lines highlighting the boundary between the allowed and ruled-out areas. As noted in the main text, there is a region of parameter space in the positronium limit with $2 \,\, {\rm MeV}\lesssim \med \lesssim 9 \,\, {\rm MeV}$ and $\qd \gtrsim 2\times10^{-8}$ unconstrained by $\Delta N_{\rm eff}$. This region corresponds to scenarios where the dark sector thermalizes with the photons at early times, but the annihilation of the dark sector fermions happens in between neutrino decoupling and electron-positron annihilation.}
    \label{fig:neff_contour}
\end{figure*}
\subsection{\label{sec:level2.1}The Dark Photon Background}
The kinetic mixing $\epsilon$ between electromagnetism and the dark sector implies the presence of an irreducible background of dark photons produced in the very early Universe. This relic abundance of dark photons is calculated by numerically solving the Boltzmann equation describing the joint evolution of the electromagnetic plasma and dark plasma energy densities:
%
\begin{align} \label{eq:boltzmannrho}
    \begin{split}
        \frac{\id \rho_{\rm \gamma}}{\id t} &= -3H(\rho_{\rm \gamma} + P_{\rm \gamma}) + C_\gamma, \\
        \frac{\id  \rho_{\rm D}}{\id t} &= -3H(\rho_{\rm D} + P_{\rm D}) +  C_{\rm D}, 
    \end{split}
\end{align} 
%
where $\rho_i$ and $P_i$ are the energy density and pressure summed over the particle species in each sector, and $C_{\rm \gamma} = -C_{\rm D}$ are the collision integrals of the energy transfer between the electromagnetic and dark sector, summed over all relevant processes. We assume that the electromagnetic plasma and dark plasma are each separately in thermal equilibrium, allowing each sector to be described by a single temperature parameter. The evolution of these temperatures is given by applying the chain rule to equation \eqref{eq:boltzmannrho} \cite{Escudero:2018mvt}
%
\begin{align} \label{eq:boltzmann}
    \begin{split}
        \frac{\id T_{\rm \gamma}}{\id t} &= \frac{-3H(\rho_{\rm \gamma} + P_{\rm \gamma}) + C_{\rm \gamma}}{\partial \rho_{\gamma}/\partial T_{\rm \gamma}}, \\
        \frac{\id T_{\rm D}}{\id t} &= \frac{-3H(\rho_{\rm D} + P_{\rm D}) + C_{\rm D}}{\partial \rho_{\rm D}/\partial T_{\rm D}}.  
    \end{split}
\end{align} 
%
For the atomic dark matter model we consider, the following energy transfer channels are relevant: annihilation of Standard Model fermions into dark fermions, Coulomb scattering between Standard Model fermions and dark fermions, and decays of plasmons and Z-bosons into dark fermions. These rates depend directly on the masses of the two dark fermions and the milli-charge $q_{\rm D}$. In principle, a Compton-like process depending directly on $\alpha_{\rm D}$ also couples the dark and visible sectors, however for $\alpha_{\rm D} \lesssim 0.06$ this process is subdominant to Coulomb scattering. We therefore ignore this process as it does not affect the dark photon yield in this regime\footnote{The relevant constraints from spectral distortions we compute in this work will also satisfy $\alpha_{\rm D} \lesssim 0.06$}.

We compute the energy transfer collision integrals following \cite{Adshead:2022ovo,Vogel:2013raa}. Starting with Equation~\eqref{eq:boltzmann}, we use initial conditions $T_{\rm \gamma,0} = 100 ~m_{\rm p_D}$ and $T_{\rm D,0} \ll T_{\rm \gamma,0}$, and evolve the system until $T_{\rm \gamma} \approx 10^{-4} ~\rm{MeV}$. We model neutrino decoupling as instantaneously occurring at a temperature of $T_{\nu, \rm dec} = 3~\rm{MeV}$. The final results of this calculation are the three temperatures $T_{\rm \gamma}$, $T_{\nu}$, and $T_{\rm D}$ allowing us to compute $N_{\rm eff}$:
%
\begin{equation} \label{eq:Neff}
    N_{\rm eff} =\frac{8}{7} \bigg( \frac{11}{4} \bigg)^{4/3}\frac{\rho_{\rm D}(T_{\rm D}) + \rho_\nu(T_\nu)}{\rho_\gamma(T_\gamma)}.
\end{equation}
%
The full details of how we numerically solve Equation~\eqref{eq:boltzmann} as well as descriptions of the relevant energy transfer processes are given in Appendix \ref{sec:appA}. 

As stated above, the rates coupling the dark sector and Standard Model, and therefore controlling the relic abundance of dark photons, depend only on the free parameters $\qd$, $\med$, and $\mpd$. We consider $10^{-10} <\qd < 10^{-4}$, $0.1~{\rm{MeV}} <\med < 100~\rm{MeV}$, and, in order to reduce the dimensionality of the parameter space, two limiting cases for $\mpd$:  the positronium limit $\mpd =  \med$ and the hydrogen limit $\mpd \gg \med$, realized here as $\mpd=1000\, \med$. In the hydrogen limit, the annihilation of the lighter fermions dominate the contribution to $N_{\rm eff}$. Figure \ref{fig:neff_contour} shows contours of the value of $N_{\rm eff}$ in the hydrogen (left) and positronium (right) limits, with the red contour demarcating the 2$\sigma$ constrained region obtained from the recent Planck+ACT+LSS+BBN analysis \cite{ACT2025}. In both limits of the model, the lower boundary between constrained and unconstrained regions of the parameter space does not improve very much with increased precision on $N_{\rm eff}$. This is because these constraints are set by a freeze-in population of dark photons, with $q_{\rm D}$ too small to thermalize the Standard Model and dark sector. $N_{\rm eff}$ therefore changes very rapidly with $q_{\rm D}$ in this region of parameter space, and we will see in \ref{sec:level3} that the projections of spectral distortion constraints from future experiments will be stronger than $N_{\rm eff}$ in this section of the parameter space.

\begin{figure}[t]
    \centering
    \includegraphics[width=1.0\linewidth]{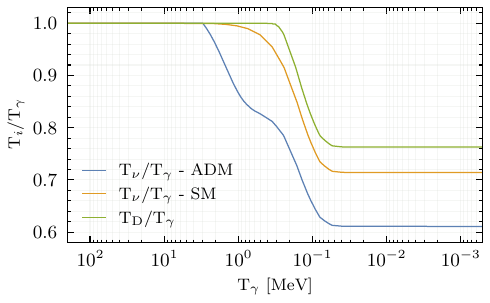}
    \caption{Evolution of the temperature ratios $T_{\nu}/T_{\gamma}$ and $T_{\rm D}/T_{\gamma}$ as a function of $T_{\gamma}$ in the positronium limit for the parameter point $\med = 5.18~\rm{MeV}$ and $\qd = 10^{-5}$, chosen to lie within the so-called `island' described in the main-text. The annihilation of dark fermions heats the photons, lowering $T_{\nu}/T_{\gamma}$ from its SM value. The dark sector does not fully decouple prior to electron-positron annihilation and so $T_{\rm D}> T^{\rm SM}_{\nu}$ at late times.}
    \label{fig:example_thermal_history}
\end{figure}

\begin{figure}[t]
    \centering
    \includegraphics[width=1.0\linewidth]{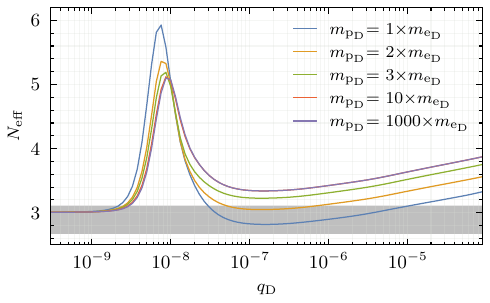}
    \caption{One dimensional slices of the $N_{\rm eff}$ contours shown in Figure~ {\ref{fig:neff_contour}} for different dark proton to dark electron mass ratios. The gray band corresponds to the 1D 95\% CL on N$_{\rm eff}$ from the most recent combined CMB and BBN analysis by the ACT collaboration \cite{ACT2025}. As the mass ratio increases, the value of $N_{\rm eff}$ quickly asymptotes to its value in the hydrogen limit.} 
    \label{fig:neff_slice}
\end{figure}

In the positronium limit, there is an `island' in the parameter space for $2\,\,{\rm MeV} \lesssim \med \lesssim 9 \,\, {\rm MeV}$ and dark electron milli-charges $\qd \gtrsim 2\times10^{-8}$ where the value of $N_{\rm eff}$ is smaller than in the surrounding area, corresponding to cases where the dark sector thermalizes with the Standard Model and the light dark fermions annihilate in between neutrino decoupling and electron-positron annihilation. The injected entropy from the dark fermion annihilations heats the photon bath relative to the neutrinos, lowering the neutrino contribution to $N_{\rm eff}$ relative to the SM. The subsequent electron-positron annihilations heat the photons relative to the dark photons. In the hydrogen limit the contribution to $N_{\rm eff}$ from the dark photons more than makes up for this suppression of the neutrino temperature and $N_{\rm eff}$ is always larger than its Standard Model value. In the positronium limit, two dark fermion species annihilate after neutrino decoupling, suppressing the neutrino temperature even further. For some values of $q_{\rm D}$, this results in a smaller value for $N_{\rm eff}$ than in the Standard Model, and as a consequence, the so-called island is not constrained by current measurements of $N_{\rm eff}$. 
A simple analysis based on conservation of entropy and instantaneous dark sector decoupling suggests a value of $N_{\rm eff} = 2.38$ in the island, however the decoupling of the dark sector overlaps partially with electron-positron annihilation and the exact value of $N_{\rm eff}$ is therefore larger. Figure~\ref{fig:example_thermal_history} shows an example thermal history for a particular choice of parameters in the positronium limit, chosen to illustrate this point. Figure~\ref{fig:neff_slice} shows the value of $N_{\rm eff}$ as a function of the milli-charge ($q_{\rm D}$) for a fixed $m_{\rm e_D} = 5.18~\rm MeV$ and various ratios of $m_{\rm p_D}/m_{\rm e_D}$, demonstrating that the value of $N_{\rm eff}$ very quickly converges to its hydrogen limit value as the mass of the dark proton increases. 
Recent determinations~\cite{Yeh:2026pil} of $N_{\rm eff}$ using new observations~\cite{Aver:2026dxv} of the primordial helium and deuterium abundances suggests $N_{\rm eff} = 2.925 \pm 0.082$ (1$\sigma$). Using this determination of $N_{\rm eff}$ instead of the recent ACT value would slightly enlarge the unconstrained island in the positronium limit due to moving the central value back towards 3. In principle the analysis presented in~\cite{Yeh:2026pil} could be combined with the ACT analysis to provide even tighter error bars on $N_{\rm eff}$.

It should be noted that the constraints on the parameter space of milli-charged particles from SN1987A presented in ~\cite{Chang:2018rso} partially overlap the island in the $N_{\rm eff}$ constraints that we compute, which we show in Fig.~\ref{fig:constraints_positronium} and Fig.~\ref{fig:constraints_hydrogen}. A recent work~\cite{Fiorillo:2024upk} on supernova bounds in self-interacting dark sectors showed that the bounds are much weaker than previously thought for dark matter masses less than 1 GeV. It is possible that the overlap of the supernova constraints and the island in $N_{\rm eff}$ is much smaller, provided their analysis holds to higher masses. 


\subsection{\label{sec:level2.2}Dark Plasmas}
Irrespective of the vacuum mass of the (dark) photon, its interactions with the charged medium will generate an effective plasma mass, leading to a wide variety of interesting phenomenological consequences. Being dependent on local quantities (such as the free electron number density), this thermal mass is dynamical and will vary both with cosmological time and length scales. In this work we will focus on the evolution of the homogeneous background mode, neglecting spatial perturbations. Patchy screening of the CMB via resonant conversions in galaxies has been considered in pure dark photon setups \cite{Pirvu2023, McCarthy2024} and provides an interesting avenue for study. We leave the extension of that mechanism in atomic dark sectors to future work.

Using a simple Drude model setup\footnote{Our derivation of the thermal mass follows closely what is found in the appendix of Berlin \textit{et al.}\cite{Berlin:2022hmt}.}, one can model the cosmological evolution of the plasma mass for both dark and visible photons. The motion of free and bound (Standard Model) electrons at position $\textbf{x}_{\rm e}$ in an external electric field ($\textbf{E}$) are described using the Euler equation
%
\begin{align}
    \partial^2_{t} \textbf{x}_{\rm e} + \Gamma\, \partial _t \textbf{x}_{\rm e} &\simeq - \frac{e}{m_{\rm e}} \textbf{E} \hspace{8mm} \textrm{(Free)}, \label{eq:SM_free}\\
    \partial^2_{t} \textbf{x}_{\rm e,b} + E_{\rm HI}^2  \,\textbf{x}_{\rm e,b} &\simeq  - \frac{e}{m_{\rm e}} \textbf{E} \hspace{8mm} \textrm{(Bound)}. \label{eq:SM-bound}
\end{align}
%
Here, $\Gamma$ is the drag term felt by the electrons and is given at lowest order by the Coulomb scattering rate with the protons. The binding energy of hydrogen sets the oscillation strength in the bound case, and is given by $E_{\rm HI} = \alpha_{\rm EM}^2 m_{\rm e}/2$. Corrections to this expression from magnetic fields enter at higher order in $\textbf{v}_{\rm e}$ thus we neglect them here. Working in Lorenz gauge ($\partial_\mu A^\mu = 0$), the electric field is related to the vector potential through $\textbf{E} = - \mathbf{\nabla} A^0 - \partial_t \textbf{A}$. 

The above expressions are valid in the limit $m_{\rm e} \ll m_{\rm p}$, which is of course satisfied in the Standard Model. In the dark sector, however, such a hierarchy of masses need not exist, such as in the positronium limit where $m_{\rm e_{D}} = m_{\rm p_{D}}$. When this is the case, the motion of free dark fermions is governed through the coupled expression
%
\begin{align}
    \partial^2_t \textbf{x}_{\rm e_{\rm D}} + \Gamma_{\rm e_{\rm D}}\, \partial _t (\textbf{x}_{\rm e_{\rm D}}-\textbf{x}_{\rm p_{\rm D}}) &\simeq - \frac{e_{\rm D}}{m_{\rm e_{\rm D}}} (\textbf{E}_{\rm D} + \epsilon\textbf{E}), \nonumber \\
    \partial^2_t \textbf{x}_{\rm p_{\rm D}} + \Gamma_{\rm p_{\rm D}}\, \partial_t (\textbf{x}_{\rm p_{\rm D}} - \textbf{x}_{\rm e_{\rm D}}) &\simeq \frac{e_{\rm D}}{m_{\rm p_{\rm D}}} (\textbf{E}_{\rm D} + \epsilon\textbf{E}), \nonumber 
\end{align}
%
where $\Gamma_{e_{\rm D}}$ and $\Gamma_{p_{\rm D}}$ are the drag forces induced due to dark Coulomb interactions on the dark electron and dark protons, respectively. These expressions can be written in analogy to Eq.~\eqref{eq:SM_free} by defining $\textbf{x}_{\rm D,rel} = \textbf{x}_{\rm e_{D}} - \textbf{x}_{\rm p_D}$, in which case we find
%
\begin{align}
    \partial^2_t \textbf{x}_{\rm D,rel} + \Gamma_{\rm D}\, \partial _t \textbf{x}_{\rm D,rel} &\simeq - \frac{e_{\rm D}}{\mu_{\rm D}} (\textbf{E}_{\rm D} + \epsilon\textbf{E}). \label{eq:DM-free}
\end{align}
%
We have introduced $\mu_{\rm D} = (m_{\rm e_{\rm D}}^{-1}+m_{\rm p_{D}}^{-1})^{-1}$ to be the reduced mass of the dark electron-proton pair, and $\Gamma_D = \Gamma_{e_{\rm D}} + \Gamma_{p_{\rm D}}$ to be the combined damping rate. For the dark hydrogen, we also find
%
\begin{align}
    \partial^2_t \textbf{x}_{\rm D,rel,b} + E_{\rm HI,D}^2  \,\textbf{x}_{\rm D,rel,b} &\simeq  - \frac{e_{\rm D}}{\mu_{\rm D}} (\textbf{E}_{\rm D} + \epsilon \textbf{E}). \label{eq:DM-bound}
\end{align}
%
For completeness, the binding energy is $E_{\rm HI,D} = \alpha_{\rm D}^2 \mu_{\rm D}/2$. Note that in the limit $m_{\rm e_{D}} \ll m_{\rm p_D}$, one has $\mu_{\rm D} \rightarrow m_{\rm e_D}$, $\textbf{x}_{\rm D,rel} \rightarrow \textbf{x}_{\rm e_D}$, and  $\Gamma_{\rm D}\rightarrow\Gamma_{e_{\rm D}}$; these two expressions then reduce cleanly to Eqs.~\eqref{eq:SM_free} and \eqref{eq:SM-bound}. The equations of motion for the SM and dark photon are given by
%
\begin{align}
    \partial^2 A_{\mu} &= j_{\mu} + \epsilon j_{\mu,\rm D}, \nonumber \\
    \partial^2 A_{\mu, \rm{D}} &= j_{\mu, \rm{D}}, \nonumber
\end{align}
%
where the spatial part of the four-current is simply given by $\textbf{j} = - e(n_{\rm e} \textbf{v}_{\rm e} + n_{\rm HI} \textbf{v}_{\rm e,b})$ with an analogous form for the dark current. Here, $n_{\rm HI}$ ($n_{\rm HI,D}$) is the number density of (dark) hydrogen. With this in hand, it is straightforward to compute the dispersion relations for the transverse and longitudinal modes in each sector \cite{Berlin:2022hmt}. In Fourier space, the equations of motion read
%
\begin{align}
    (\omega^2 - k^2)
    \begin{pmatrix}
    \tilde{\textbf{A}}_{\rm T} \\
    \tilde{\textbf{A}}_{\rm T,\rm{D}}
    \end{pmatrix}
    &\simeq
    \begin{pmatrix}
    m_{\gamma}^2 + \epsilon^2 m_{\gamma_{\rm D}}^{\,2}
    &
    \epsilon m_{\gamma_{\rm D}}^{\,2}
    \\[6pt]
    \epsilon m_{\gamma_{\rm D}}^{\,2}
    &
    m_{\gamma_{\rm D}}^{\,2}
    \end{pmatrix}
    \begin{pmatrix}
    \tilde{\textbf{A}}_{\rm T} \\
    \tilde{\textbf{A}}_{\rm T,\rm{D}}
    \end{pmatrix},
    \\[12pt]
    \omega^2
    \begin{pmatrix}
    \tilde{\textbf{A}}_{\rm L} \\
    \tilde{\textbf{A}}_{\rm L,\rm{D}}
    \end{pmatrix}
    &\simeq
    \begin{pmatrix}
   m_{\gamma}^2 + \epsilon^2 m_{\gamma_{\rm D}}^{\,2}
    &
    \epsilon m_{\gamma_{\rm D}}^{\,2}
    \\[6pt]
    \epsilon m_{\gamma_{\rm D}}^{\,2}
    &
    m_{\gamma_{\rm D}}^{\,2}
    \end{pmatrix}
    \begin{pmatrix}
    \tilde{\textbf{A}}_{\rm L} \\
    \tilde{\textbf{A}}_{\rm L,\rm{D}}
    \end{pmatrix}.
\end{align}
%
Explicitly, the effective mass term for the photon is given by 
%
\begin{align}
    m_{\gamma}^2 &\simeq \frac{e^2 n_{\rm e}}{m_{\rm e}}\left(\frac{1}{1 + i \Gamma/\omega} - \frac{n_{\rm HI}}{n_{\rm e}} \frac{1}{(E_{\rm HI}/\omega)^2-1} \right), \nonumber\\
    &\simeq  \frac{e^2 n_{\rm e}}{m_{\rm e}}\left(1- 5.4 \times 10^{-3}\frac{n_{\rm HI}}{n_{\rm e}} \left[\frac{\omega}{{\rm eV}}\right]^2\right), \label{eq:plasma_mass}
\end{align}
%
where in the last line we have evaluated $E_{\rm HI}$, and have also assumed the hierarchy $\Gamma \ll \omega \ll E_{\rm HI}$. Indeed, For the redshifts of conversion we will be interested in $10^3 \leq z \leq 2\times 10^6$, the (visible) Coulomb scattering rate is extremely small such that $\Gamma/\omega_{\rm max} \ll 1$ where $\omega_{\rm max}$ is the peak CMB frequency at a given redshift\cite{Berlin:2022hmt}. In principle, pre-recombination (in the visible sector), conversions will take place with photons satisfying $E_{\rm HI}/\omega <1$. However, at $z \gtrsim 10^3$, the hydrogen fraction is heavily suppressed such that $n_{\rm HI}/n_{\rm e} \ll 1$, thus nullifying the contribution from neutral hydrogen to the thermal mass. 

The above expression is valid at leading order for a plasma of non-relativistic electrons, as is the case during our redshifts of interest \cite{Chluba2024}. We neglect the correction to the mass coming from free protons, as their contribution scales as $m_{\gamma,{\rm p}}^2\propto e^2 n_{\rm e}/m_{\rm p}$, a roughly two order of magnitude suppression to the plasma mass when compared to the electrons. Additionally, between $10^3 \lesssim z \lesssim 10^4$, we also neglect contributions from both HeI and HeII. The reason for this is twofold: first, the relative number of helium nuclei is suppressed relative to the electrons, $(n_{\rm HeI}+n_{\rm HeII})/n_{\rm e} \lesssim 0.1$. Second, the ionization energy of each helium state is greater than hydrogen ($E_{\rm HI} < E_{\rm HeI}, E_{\rm HeII}$), leading to a less impactful contribution relative to hydrogen. As we shall see, the resonant conversion probability is higher at low frequencies ($\omega \lesssim \omega_{\rm peak}$), which also reduces the relative importance of the helium contribution to $m_{\gamma}$. In practice, we use Eq.~\eqref{eq:plasma_mass} to evaluate the photon plasma mass, though at these early times this can be well approximated using solely the electron contribution.

Similarly, the dark photon thermal mass is given by 
%
\begin{align}
    m_{\gamma_{\rm D}}^2 \simeq e_{\rm D}^2\bigg(\left[\frac{n_{\rm e_{\rm D}}}{m_{\rm e_{\rm D}}} + \frac{n_{\rm p_{\rm D}}}{m_{\rm p_{\rm D}}} \right]&\left[\frac{1}{(1 + i \Gamma_{\rm D}/\omega)}\right] \nonumber \\
    &- \frac{n_{\rm HI,D}}{\mu_{\rm D}} \frac{1}{(E_{\rm{HI},{\rm D}}/\omega)^2-1} \bigg). 
\end{align}
%
This time, we must include contributions from the dark protons which will be relevant in the positronium limit. Dark baryogenesis ensures $n_{\rm e_{\rm D}} \simeq n_{\rm p_{\rm D}}$, allowing for some simplification of the expression. With this in mind, our expression becomes
%
\begin{align}
\begin{split}
    m_{\gamma_{\rm D}}^2 \simeq  \frac{e_{\rm D}^2 n_{\rm e_{\rm D}}}{\mu_{\rm D}}\bigg(1 
    &- \frac{n_{\rm HI, D}}{n_{\rm e_{\rm D}}} \left[\frac{{\rm eV}}{\alpha_{\rm D}^2 \mu_{\rm D}/2} \right]^{2}\left[\frac{\omega}{{\rm eV}}\right]^2 \bigg). \label{eq:dark_plasma_mass}
\end{split}
\end{align}
%
Similar to the Standard Model case, this expression is valid in the limit of $\Gamma_{\rm D}/\omega \ll 1$. Parametric estimates of this damping rate indicate that this limit is easily satisfied for the parameter space of interest to the CMB spectral distortions bounds computed below. In our setup, resonant conversions occur as the dark atomic sector is undergoing its recombination. For the parameter space we consider, the peak frequency of the dark CMB obeys $\omega_{\rm peak, D} \lesssim E_{\rm HI,D}$, and thus the vast majority of photons undergoing the conversion will also satisfy $\omega/E_{\rm HI,D} \ll 1$, justifying our treatment of the contribution of dark hydrogen in the above expression.

\subsection{\label{sec:level2.3}Dark Recombination}
The redshift evolution of the dark photon mass is governed by changes in the free dark electron (and dark proton) number density. At temperatures above the binding energy of dark hydrogen ($E_{\rm HI,D}$), the atomic dark sector is a fully ionized plasma, with free dark electron fraction $X_{\rm e_{D}}=1$. When the dark photon temperature falls below $E_{\rm HI,D}$, it becomes energetically preferable for dark electrons and dark protons to recombine into dark hydrogen, and the dark ionization fraction begins to fall exponentially in Saha equilibrium, described by the Saha equation:
%
\begin{equation}
    \frac{X_{\rm e_{D}}^{2}}{1-X_{\rm e_{D}}} = \frac{1}{n_{\rm D}}\left(\frac{T_{\rm DM}\mud}{2\pi} \right)^{3/2}e^{-E_{\rm HI,D}/T_{\rm DM}}.
\end{equation}
%
The total (free and bound) dark electron number density is $n_{\rm D}(= n_{\rm e_{\rm D}}+ n_{\rm HI,D})$, and the temperature of the atomic dark matter is $T_{\rm DM}$. At high redshifts, dark Compton interactions keep the dark photons and electrons at the same temperature ($T_{\rm D} \simeq T_{\rm DM}$). The free dark electron number density drops exponentially during dark recombination, with this process occurring at $T_{\rm D} \simeq \mathcal{O}(10^{-2}) \times E_{\rm HI,D}$ in typical atomic dark matter scenarios. From Eq.~\eqref{eq:dark_plasma_mass}, it is clear that the exponential decrease in $n_{\rm e_{D}}$ causes a rapid decline in the dark photon plasma mass. During the primordial distortion epoch ($10^3 \, \lesssim z \lesssim \, 10^6$), the SM photon mass evolves steadily as $\mg \propto (1+z)^{3}$. Thus, if dark recombination occurs in this epoch, it will generally be accompanied by a resonant crossing when $m_{\gamma} = m_{\gamma_{\rm D}}$.

By explicitly computing the dark ionization history with a modified version of the Boltzmann code CLASS for a few test points in parameter space, we confirm that this crossing generically occurs while the dark electron ionization fraction $X_{\rm e_{D}}$ is in Saha equilibrium~\cite{Lesgourgues:2011re,Blas:2011rf,Bansal:2022qbi}. This allows us to use the Saha equation to compute $X_{\rm e_{D}}$ as a function of temperature, without requiring CLASS to be run for every parameter point. Deviations from Saha equilibrium would only serve to slow the decrease of $n_{\rm e_{D}}$ with time, lengthening the time the two photon masses spend in the resonance band, and ultimately enhancing the probability of conversion. Thus, our usage of the Saha equilibrium value of $X_{\rm e_{D}}$ is a conservative assumption. 

In the positronium limit where $\mpd=\med$, one can roughly estimate the redshift of the resonance,
%
\begin{equation}
\zres \approx 
    1220 \left(\frac{\mud}{\mu_{\textrm{SM}}}\right)^{1.06}\left(\frac{\alphad}{\alpha_{\textrm{SM}}}\right)^{2.04}\left(\xid\right)^{-1.11}. \label{eq:z_res_approx}
 \end{equation}
%
Here $\mu_{\textrm{SM}}$ and $\alpha_{\textrm{SM}}$ are the SM reduced mass and fine structure constant, respectively. This scaling is accurate at the 5\% level for $\fd$ from 0.01 to 1. In the hydrogen limit, $\zres$ is also insensitive to $\mpd$. Notably, this scaling is very close to  $z_{\rm res} \sim E_{\rm HI,D}/\xid$.

\begin{figure}[t]
    \centering
    \includegraphics[width=1.0\linewidth]{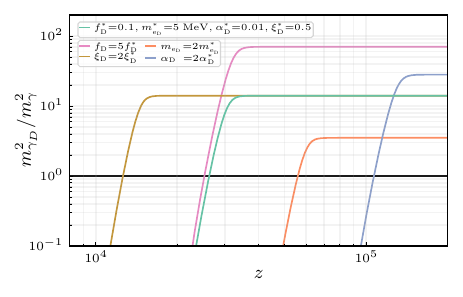}
    \caption{Ratio of dark photon plasma mass squared to SM photon plasma mass squared as a function of redshift, for several choices of aDM parameters. The dark proton mass $\mpd$ is set equal to $\med$ in all cases (the positronium limit). The resonant redshift is determined by finding where a given contour crosses the $(m_{\gamma_{\rm D}}/m_{\gamma})^2 = 1$ line.}
    \label{fig:photonmasses} 
\end{figure}

If the dark photon mass is lower than the SM photon mass before dark hydrogen recombination, there is no crossing. The condition for the dark photon plasma mass to be greater than the SM photon plasma mass at times before both SM and dark recombination, thus allowing a resonance to occur during dark recombination, is
%
\begin{equation}
    \left(\frac{\alphad}{\alpha_{\textrm{SM}}}\right)\left(\frac{\mu_{\textrm{SM}}}{\mud} \right)\left(\frac{m_{H}}{\mHd} \right)\left( \frac{\fd\Omega_{DM}}{\Omega_{b}}\right) > 1.
\end{equation}
%
We showcase the dependence of the evolution of $\mgd$ on the various model parameters in Figure~\ref{fig:photonmasses}. In this plot, the teal contour provides a rather generic point in parameter space that we take as a benchmark. Each of the other contours modify one parameter relative to this benchmark, illustrating the effect on the resonant redshift. Raising $\fd$ (pink) increases the dark photon plasma mass but does not significantly change the redshift of the dark recombination or crossing with the SM photon mass. Boosting $\med$ (orange) increases the redshift of the dark recombination, but decreases the dark photon plasma mass prior to the dark recombination. Increasing $\alphad$ (blue) causes the redshift of the dark recombination to happen at earlier times and raises the dark photon plasma mass. Finally, increasing the dark photon temperature (gold) decreases the redshift of dark recombination but does not affect the dark photon mass. 
%
%

\section{\label{sec:level3}Spectral Distortions from Resonant Conversions}
\subsection{\label{sec:level3.1}CMB Spectral Distortion Formalism}
As discussed above, dark photons minimally couple to the Standard Model through a kinetic mixing interaction. This kinetic mixing allows for the possibility of both perturbative  as well as resonant conversions, the latter of which will be the main study in this section. Resonant conversions provide an abrupt transfer of energy and entropy both into and out of the dark sector, localized in time around the redshift at which the condition $m_{\gamma} \simeq m_{\gamma_{\rm D}}$ is satisfied. Generally speaking, when the resonant condition is satisfied in the CMB distortion epoch, the perturbative conversion scenario can safely be neglected \cite{McDermott2019}.

Spectral distortions in scenarios with dark photons have been considered numerous times in the literature recently \cite{McDermott2019, Berlin:2022hmt, Chluba2024, Arsenadze2024}. However, these constraints often rely on the assumption that the dark photon is an isolated dark particle, comprising of either all of, or none of, the dark matter content in the universe. In our scenario, the dark photon possesses dark electromagnetic interactions with both a dark electron and proton, imbuing it with its own plasma mass and allowing the formation of a dark CMB. The contribution of this dark CMB to $\Delta N_{\rm eff}$ was discussed in Sec.~\ref{sec:level2.1}.

Berlin \textit{et al}. \cite{Berlin:2022hmt} have studied a similar situation in which their dark sector consisted of a photon and electron. This endows the dark photon with a plasma mass, scaling with $m_{\gamma_{\rm D}} \sim (1+z)^{3/2}$ at all redshifts for non-relativistic dark electrons, ensuring that the resonance condition is only satisfied post-(visible) recombination. While they showed it was possible to put leading constraints on (vacuum) massless dark photons from spectral distortions in their setup, this post-recombination restriction did not allow them to take full advantage of current constraints, as it was not possible to probe pre-recombination conversions. Such a scenario can only be realized when the dark plasma mass undergoes a rapid change, such as what occurs during dark recombination.

For the purposes of monopole (sky-averaged) CMB spectral distortions, there are effectively three methods that can be utilized to predict the expected signal from exotic sources. At high redshifts ($5 \times 10^4 \lesssim z \lesssim 2 \times 10^6$), the injection and extraction of energy/entropy will induce a non-zero chemical potential ($\mu$) in the plasma, due to the inefficiency of number changing interactions (double Compton and Bremsstrahlung) \cite{Zeldovich1969, Sunyaev1970SPEC, Burigana1991, Hu1993, Chluba2011therm, Sunyaev2013, Tashiro2014, Lucca2020}. For these primordial $\mu$-distortions, it has been shown \cite{Chluba2013, Chluba2015} that a reasonable approximation to the full numerical solution is given by 
%
\begin{align} \label{eq:mu-greens-function}
	\mu \simeq  1.401 \int \id z \left[ \frac{1}{\rho_{\rm CMB}} \frac{\id \rho_{\rm inj}}{\id z} - \frac{4}{3 N_{\rm CMB}} \frac{\id N_{\rm inj}}{\id z} \right]\mathcal{J_{\mu}}(z).
\end{align}
%
Where $\id \rho_{\rm inj}/\id z$ and $\id N_{\rm inj}/\id z$ are the rates of energy/entropy injection or extraction into the CMB as a function of redshift, and $\mathcal{J}_{\mu}(z)$ is the distortion visibility function, where we use Method C as presented in \cite{Cyr2023}. The 2$\sigma$ bounds from COBE/FIRAS \cite{Fixsen1996, Fixsen2003, Bianchini2022} are $\mu < 4.7 \times 10^{-5}$, while forecasts from the currently proposed FOSSIL experiment are much stronger, at the $\mu \lesssim 10^{-8}$ level.

As one transitions to lower (but still pre-recombination, $z \lesssim 5 \times 10^4$) redshifts, Compton scattering becomes inefficient at mediating energy redistribution across the spectrum. This usually marks the onset of the so-called $y$-distortion epoch, and for direct photon injection/extraction scenarios, it often becomes necessary to go beyond the simple prescription given above and utilize a numerical framework to compute the evolution of the photon phase space. For this purpose, both \texttt{CosmoTherm} \cite{Chluba2012} (low energy injections, $E_{\rm inj} \lesssim T_{\gamma}$) and \texttt{DarkHistory} \cite{Liu2019, Liu2023a, Liu2023b} (high energy injections $E_{\rm inj} \gtrsim T_{\gamma}$) are good choices. 

For resonant dark photon conversions, the energy transfer typically takes place in the low energy regime. When this is the case, the total pre-recombination effective energy release ($\Delta \rho_{\rm eff}/\rho$) has been shown to be a reasonable approximation to the constraints derived by considering the full numerical solution as computed\footnote{The full numerical implementation was performed in \cite{Chluba2024} for the vanilla dark photon scenario.} in \texttt{CosmoTherm} \cite{Chluba2024}. This is given by a slight variation on Eq.~\eqref{eq:mu-greens-function}, namely
%
\begin{align} \label{eq:eff-energy}
	\frac{\Delta \rho_{\rm eff}}{\rho} \simeq   \int \id z \left[ \frac{1}{\rho_{\rm CMB}} \frac{\id \rho_{\rm inj}}{\id z} - \frac{4}{3 N_{\rm CMB}} \frac{\id N_{\rm inj}}{\id z} \right]\mathcal{J_{\rm eff}}(z),
\end{align}
%
\begin{figure*}[t]
    \centering
    \includegraphics[width=1.0\textwidth]{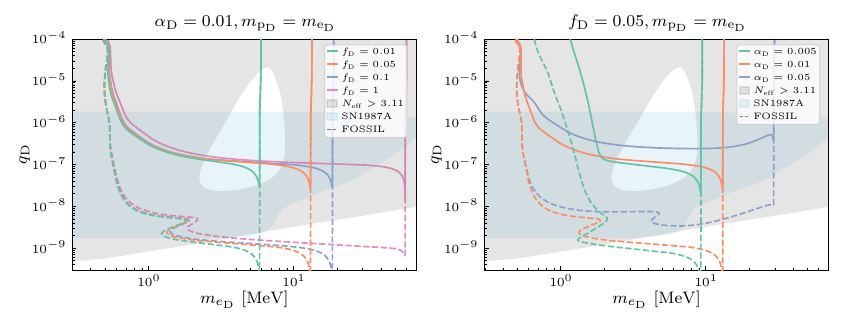}
    \caption{Constraints on $\qd$ as a function of $\med$ in the positronium limit ($\mpd=\med$). Left: constraints with $\alphad=0.01$. Right: constraints with $\fd=0.05$. We show current constraints from COBE/FIRAS in solid and projected constraints from FOSSIL in dashed. The constraint from $N_{\mathrm{eff}}$ as measured by ACT is shaded gray, and the constraint from SN1987A is shaded blue. }
    \label{fig:constraints_positronium}
\end{figure*}
where we use $\mathcal{J_{\rm eff}} = {\rm e}^{-(z/z_{\rm th})^{5/2}}/(1 + {\rm e}^{-20(z-z_{\rm rec})/z_{\rm rec}})$ with $z_{\rm th}=2\times10^{6}$ and $z_{\rm rec}=1300$ as a smooth window function which selects only the pre-recombination part of the injection. At $2\sigma$, the COBE/FIRAS constraint on the effective energy injection is $\Delta \rho_{\rm eff}/\rho \leq 3 \times 10^{-5}$.

Post-recombination, it is often assumed that the distortion is completely frozen-in, and it is therefore best practice to compare with the residuals measured by the COBE/FIRAS experiment. Indeed, this is what was done by Berlin \textit{et al.}, and is reasonable under the assumption that distortions are formed primarily within the COBE/FIRAS frequency bands ($\simeq$ 60 GHz - 600 GHz). If distortions are broadband, the thermal state of the Universe should in principle be evolved alongside the distortion to ensure adverse effects are not produced elsewhere. Specifically, the injection of high energy photons post recombination can lead to particle cascades, and injecting low frequency photons can initiate the soft photon heating effect, leading to potential signatures in 21cm cosmology \cite{Acharya2023, Cyr2024}. We leave an in-depth analysis of these effects to future work, and instead choose to focus solely on distortions stemming from pre-recombination conversions.

To determine the initial distortion, we follow the usual Landau-Zener approach\footnote{Whether the resonance condition can be maintained in realistic cosmological environments for (vacuum) massive dark photons has been recently been brought into question \cite{Hook2025}. We argue in App.~\ref{sec:appC} that the massless dark photon suffers no such issues.} which describes the frequency dependent conversion probability for the $\gamma \leftrightarrow \gamma_{\rm D}$ process on resonance. For a given set of atomic dark matter parameters, the conversion redshift can be found by matching Eqs.~\eqref{eq:plasma_mass} and \eqref{eq:dark_plasma_mass}. The conversion probability at the resonance is given by \cite{Berlin:2022hmt}
%
\begin{align}
    P_{\gamma \leftrightarrow \gamma_{\rm D}}(x) = 1 - {\rm exp}\bigg( - &\frac{\pi \epsilon^2}{x \, T_{\rm 0}} \sum_{\rm res} \frac{m_{\gamma_{\rm D}}^2}{H(1+z)^2} \nonumber \\
    &\times \left| \frac{\id \log m_{\gamma}^2}{\id z} - \frac{\id \log m_{\gamma_{\rm D}}^2}{\id z} \right|^{-1} \bigg),
    \label{eq:landau-zener}
\end{align}
%
where it is understood that each of the quantities are evaluated at all resonance redshifts. We label photon frequencies in a dimensionless form through $x = \omega/T_{\gamma}(z)$, where at all times the peak of the blackbody spectrum lies at $x \simeq 2$. In the above, $T_0$ is the CMB temperature today and $H$ the Hubble rate at the resonance. It should be noted that the conversion probability is frequency dependent, with low frequency photons converting more readily between the two sectors. In a conversion, a dark photon with $\omega_{\rm d}$ will convert to a photon of the same frequency (and vice-versa). When both the standard and dark photon sectors are populated, the distortion induced in the Standard Model spectrum (assuming one resonance) can be written
%
\begin{align} \label{eq:delta_n}
    \Delta n_{\gamma}(x) &= - n_{\gamma} P_{\gamma \rightarrow \gamma_{\rm D}} + n_{\rm d} P_{\gamma_{\rm D} \rightarrow \gamma}, \nonumber \\
    &= P_{\gamma \leftrightarrow \gamma_{\rm D}} \left( \frac{1}{{\rm e}^{x (T_{\gamma}/T_{\rm D})}-1} - \frac{1}{{\rm e}^{x}-1} \right).
\end{align}
%
Here we have assumed that both the standard and dark radiation backgrounds are initially free from any distortions, e.g. they start out as independent blackbodies with occupation numbers $n_{i} = ({\rm e}^{x_i}-1)^{-1}$. In the case of a pre-recombination $\delta$-function injection, the source terms are simply given by
%
\begin{align}
    \label{eq:drhodz}
    \frac{\id \rho_{\rm inj}}{\id z} &=   \frac{T^4_{\gamma}(z_{\rm res})}{\pi^2} \int \id x \,x^3 \Delta n_{\gamma}(x)  \delta (z - z_{\rm res}), \\
    \frac{\id N_{\rm inj}}{\id z} &=  \frac{T^3_{\gamma}(z_{\rm res})}{\pi^2} \int \id x \,x^2 \Delta n_{\gamma}(x) \delta (z - z_{\rm res}).
\end{align}
%
From Eq.~\eqref{eq:delta_n} it is clear that for $T_{\rm D} \lesssim T_{\gamma}$, the direction of overall energy transfer is into the dark sector. 

\subsection{\label{sec:level3.2}CMB Spectral Distortion Limits and Projections}

With the formalism described above, we are equipped to compute the effective energy release caused by resonant conversion between the dark photons and SM photons at the time of dark hydrogen recombination. We utilize the $2\sigma$ bound of $\Delta \rho_{\rm eff}/\rho \leq 3 \times 10^{-5}$ from COBE/FIRAS \cite{Chluba2024}, and also project constraints from the proposed FOSSIL experiment, using $\Delta \rho_{\rm eff}/\rho=10^{-8}$ as the forecasted limit.
 
\begin{figure*}[t]
    \centering
    \includegraphics[width=1.0\linewidth]{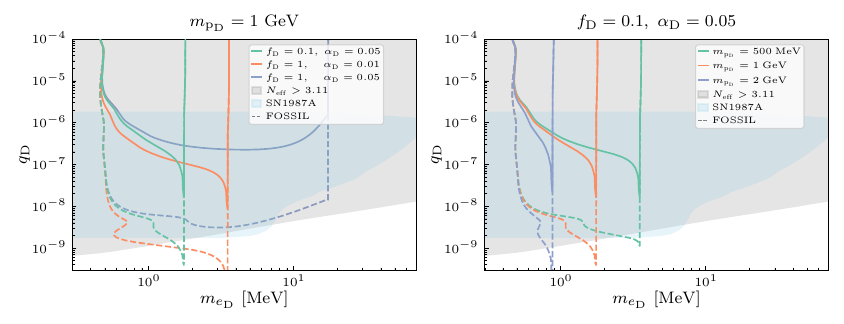}
    \caption{Constraints on $\qd$ as a function of $\med$ in the hydrogen limit ($\mpd\gg\med$). Left: constraints with $\mpd=1$ GeV. Right: constraints with $\fd=0.1$ and $\alphad=0.05$. We show current constraints from COBE/FIRAS in solid and projected constraints from FOSSIL in dashed. The constraint from $N_{\mathrm{eff}}$ as measured by ACT is shaded in gray, and the constraint from SN1987A is shaded blue. Lower values of $\fd$ and $\alphad$ rapidly become unconstrained in this parameter space. }
    \label{fig:constraints_hydrogen}
\end{figure*}

In Figure~\ref{fig:constraints_positronium} we show constraints in the space of $\med$ and $\qd$ for fixed values of $\fd$ and $\alphad$ in the positronium limit. The left edge of these constraints is set by the redshift of the resonance falling below that of SM recombination. As $\alphad$ increases, this left edge moves to lower $\med$ values due to the scaling described in Eq.~\eqref{eq:z_res_approx}. The right edge is set by the dark photon mass falling below the SM photon mass at high redshifts, eliminating the possibility of a resonance entirely. Prior to dark recombination, $m^2_{\rm \gamma_D} \propto \alpha_D f_{\rm D}/(m_{\rm{e_D}}\mpd)$ and the right edge of the limits therefore move to larger $m_{\rm e_D}$ with increasing $\fd$ or $\alphad$.

Away from the boundaries, the constraint on the dark charge is relatively flat with respect to $m_{\rm {e_D}}$ and $f_{\rm D}$. Depending on the choice of the dark gauge coupling $\alpha_{\rm D}$, the upper bound on the milli-charge is typically in the range of $q_{\rm D}\lesssim 10^{-7}-10^{-6}$. Near the right edge of the constraints however, the bound on $q_{\rm D}$ starts to rapidly get stronger before being cut off entirely due to the early time dark photon plasma mass going below the visible photon plasma mass. This is caused by the fact that the dark photon plasma mass decreases exponentially with redshift during dark recombination. From Eq. \ref{eq:landau-zener} we see that the conversion probability depends on the inverse difference of the slopes of the two plasma masses, which will typically be dominated by the rapidly decreasing dark photon thermal mass. For level crossings happening right at the onset of dark recombination, the two slopes will be similar in magnitude and the conversion probability will thus be larger. When considering forecasted limits from the FOSSIL satellite, we see that the constraints can be improved by a little more than two orders of magnitude. A future space mission such as this would thus be able to push down into the low $q_{\rm D}$ region of parameter space, superseding $N_{\rm eff}$ and supernova constraints. 

In the hydrogen limit, the unconstrained island from $N_{\rm eff}$ constraints closes. Additionally, with $\mpd \gg \med$, the total dark electron number density (and therefore the dark photon mass) is now set by $f_{\rm D}/\mpd$. Thus, for larger dark proton masses, larger $\fd$ and $\alphad$ are also required to ensure that a crossing still occurs during the distortion epoch. We highlight a subset of the hydrogen limit constraints by fixing $\mpd = 1$ GeV in Figure~\ref{fig:constraints_hydrogen}. The limits from COBE/FIRAS remain roughly similar in amplitude to the positronium limit, however, no currently unconstrained regions of parameter space can be ruled out using current spectral distortion data. A measurement from FOSSIL would allow us to set novel constraints in the low charge limit.

\section{\label{sec:level4}Conclusions}

In this work, we have derived constraints on atomic dark sectors with a kinetically mixed massless dark photon, leveraging complementary bounds from $\Delta N_{\rm eff}$ and CMB spectral distortions. Our analysis extends prior studies of milli-charged dark sectors~\cite{Vogel:2013raa, Adshead:2022ovo} to include bound state formation in the dark sector, and generalizes the spectral distortion constraints of Berlin \textit{et al}.~\cite{Berlin:2022hmt} to scenarios in which the dark plasma undergoes its own recombination at redshifts $10^3 \lesssim z \lesssim 10^6$.

We first computed the irreducible dark photon abundance produced through freeze-in, solving the coupled Boltzmann equations describing energy transfer between the visible and dark sectors. Using the most recent ACT~DR6 measurement of $N_{\rm eff}$ in combination with Planck, large scale structure, and BBN, we updated constraints on the milli-charge $q_{\rm D}$ and dark electron mass $m_{\rm e_D}$ in both the hydrogen ($m_{\rm p_D} \gg m_{\rm e_D}$) and positronium ($m_{\rm p_D} = m_{\rm e_D}$) limits. In the positronium limit, we identified a region of parameter space with $2~{\rm MeV} \lesssim m_{\rm e_D} \lesssim 9~{\rm MeV}$ and $q_{\rm D} \gtrsim 2 \times 10^{-8}$ that evades the $\Delta N_{\rm eff}$ bound due to the annihilation of light dark fermions between neutrino decoupling and electron-positron annihilation.

The central result of this paper is the demonstration that resonant conversions between visible and dark photons at the epoch of dark hydrogen recombination can generate observable spectral distortions. This conversion is possible due to the exponential decrease in the free dark electron fraction during dark recombination, which causes the dark photon plasma mass to drop rapidly, allowing it to cross the SM photon plasma mass at redshifts within the primordial distortion epoch. This is in contrast to scenarios without bound state formation, where the dark and Standard Model photon masses obey the same scaling ($m_{\gamma}\propto (1+z)^{3/2}$) at all pre-recombination redshifts, restricting resonances to occur only after (or~during) visible recombination.

Using the Landau-Zener formalism to compute the conversion probability at resonance, we showed that COBE/FIRAS constrains the milli-charge at the level of $\qd \lesssim 10^{-7} - 10^{-6}$ across a broad range of dark electron masses, dark matter fractions, and dark fine structure constants. These constraints are robust: they are largely insensitive to $\med$ and $\fd$, and remain stable for $\alphad \lesssim 0.01$. Moreover, in the positronium limit, spectral distortions provide the leading constraints in the region of parameter space that is unconstrained by $\Delta N_{\rm eff}$ and SN1987A, highlighting the complementarity between these three observational channels. We additionally showed that the projected sensitivity of the proposed FOSSIL experiment could strengthen these bounds by roughly two orders of magnitude, reaching $\qd \lesssim 10^{-9} - 10^{-8}$, thus providing a clear observational target for next-generation spectral distortion missions including TMS, BISOU, and COSMO.

Several avenues for future investigation remain. First, post-recombination resonant conversions may produce broadband distortions that extend below the COBE/FIRAS frequency range, potentially sourcing signatures accessible through 21\,cm experiments. Second, the patchy screening of the CMB through resonant conversions in collapsed structures, which has been considered for vanilla dark photon models, could be extended to atomic dark sectors and may yield spatially varying spectral distortion signals. Finally, departures from Saha equilibrium during dark recombination would slow the evolution of the dark ionization fraction, lengthening the time the two plasma masses spend near resonance and enhancing the conversion probability. Our use of the Saha equation is therefore conservative, and a full treatment using a dark recombination code could modestly strengthen the constraints presented here.

\section{Acknowledgments}
The authors would like to thank Sandip Roy for contributions to the early stages of this project. Additionally, we would like to thank Uddipan Banik, Pranjal Ralegankar, Rouven Essig, Jens Chluba, and Cara Giovanetti for useful discussions. We would like to thank the referee for their close and careful reading of this work and the helpful feedback they provided.
DA is supported by DOE Grant DE-SC0025309 and Simons Investigator in Physics Awards 623940 and MPS-SIP-0001046.
BC is grateful for support from an NSERC Banting Fellowship, as well as the Simons Foundation (Grant Number 929255). JB acknowledges support from NSF grants PHY-2210533 and PHY-2513893. 
The authors would like to thank Stony Brook Research Computing and Cyberinfrastructure and the Institute for Advanced Computational Science at Stony Brook University for access to the high-performance SeaWulf computing system, which was made possible by \$1.85M in grants from the National Science Foundation (awards 1531492 and 2215987) and matching funds from the Empire State Development’s Division of Science, Technology and Innovation (NYSTAR) program (contract C210148).
\onecolumngrid
\appendix
\section{\label{sec:appA}Solving the Boltzmann Equation}

In this appendix we provide the technical details regarding our numerical implementation of the Boltzmann equation, as well as our calculations of the collision integrals for the relevant energy transfer processes. Our methodology is based on \cite{Adshead:2022ovo}, which we follow closely. Our starting point is Eq.~\eqref{eq:boltzmann}, which directly describes the time evolution of the temperatures of the Standard Model plasma and the dark sector. We take the energy density and pressure of each sector as:
%
\begin{align}
    \rho_i(T_i) = \frac{\pi^2}{30}~g_{i*,\rho}(T_i) ~T_i^4,\quad P_i = w_i(T_i)~\rho_i(T_i),
\end{align}
%
where $w_i = g_{i*,P}/3 g_{i*,\rho}$ is the equation of state of the sector i. For the Standard Model plasma we use the $g_{\rm SM*}$ from \cite{Husdal:2016haj} (with the neutrino contribution removed below the neutrino decoupling temperature), and for the dark sector we take:
%
\begin{align}
\begin{split}
    g_{D*,\rho}(T_{\rm D}) &= 2 + 4\times\frac{30}{\pi^2 T^4_{\rm D}} \int \frac{\id^3\vec p}{(2\pi)^3} \bigg [\frac{E_{e_{\rm D}}(\vec p)}{e^{E_{e_{\rm D}}(\vec p)/T_{\rm D}} +1} + \frac{E_{p_{\rm D}}(\vec p)}{e^{E_{p_{\rm D}}(\vec p)/T_{\rm D}} +1} \bigg], \\
     g_{D*,P}(T_{\rm D}) &= 2 + 4\times\frac{90}{\pi^2 T^4_{\rm D}} \int \frac{\id^3\vec p}{(2\pi)^3} \bigg [\frac{\vec p^2}{3E_{e_{\rm D}}(\vec p)}\frac{1}{e^{E_{e_{\rm D}}(\vec p)/T_{\rm D}} +1} + \frac{\vec p^2}{3E_{p_{\rm D}}(\vec p)}\frac{1}{e^{E_{p_{\rm D}}(\vec p)/T_{\rm D}} +1} \bigg].
\end{split}
\end{align}
%
The numerical integration of equation~\eqref{eq:boltzmann} is performed with the solve\textunderscore ivp method from the \texttt{integrate} module of \texttt{SciPy}~\cite{2020SciPy-NMeth}. We configure the solver to use the `BDF' method with absolute and relative error tolerances set to $10^{-8}$. As stated in the main text, we instantaneously decouple the neutrinos at a temperature of $T_{\nu,\rm dec} = 3~\rm{MeV}$. We implement this in code by first solving Eq.~\eqref{eq:boltzmann} until the Standard Model plasma reaches $T_{\nu,\rm dec}$, at which point we stop the solver and use its current state as the initial conditions for a second solver that treats the neutrinos as a separate, decoupled species undergoing adiabatic cooling. Our treatment of the neutrinos yields a Standard Model result $N_{\rm eff} = 3.009$, slightly lower than state-of-the-art calculations of $N^{\rm SotA}_{\rm eff} = 3.044$ \cite{Escudero:2018mvt,EscuderoAbenza:2020cmq,Cielo:2023bqp,Bennett:2019ewm,Akita:2020szl,Froustey:2020mcq,Bennett:2020zkv,Akita:2022hlx}. For each ADM parameter point, we compute $\Delta N_{\rm eff}$ by subtracting our Standard Model value from the BSM value we compute using Eq.~\eqref{eq:Neff} with the final temperatures of the photons, dark photons, and neutrinos.  
 
For the atomic dark matter model described in Sec.~\ref{sec:level2}, the relevant energy transfer channels are: fermion annihilation, Coulomb scattering, plasmon decay, and Z-boson decay. Above the QCD phase transition temperature, which we take as $T_{\rm QCD} = 200~\rm{MeV}$, we include the contributions of the quarks to the total annihilation and scattering rates as well as the photon plasma mass. Above the electroweak phase transition, which we take to happen at $T_{\rm EW} = 160~\rm{GeV}$, Coulomb scattering is mediated by the hyper-charge gauge boson and includes contributions from the Higgs doublet and neutrinos.
 
In the following four subsections we summarize the methods we use to simplify and compute the collision integrals and give explicit expressions for the cross sections and decay widths of the relevant processes. For more thorough derivations of these formulae, one should check Appendix C of \cite{Adshead:2022ovo}. 

\subsection{Fermion Annihilation}
In general, the forward collision term for particle-antiparticle annihilation of a Standard Model species $a$ into a dark species $b$ is given by:
%
\begin{align}
    C^{\rm ann}_{\rm F} = \int \id P \,\, S\, |\mathcal M_{a\bar a \rightarrow b \bar b}|^2(E_1 + E_2)
    \times f_1 f_2(1\pm f_3)(1\pm f_4),
\end{align}
%
with
%
\begin{equation}
    \id P = (2\pi)^4\delta^4(p_1 +p_2 -p_3-p_4)\prod_{i=0}^4 \frac{d^3 \vec{p}_i}{(2\pi)^3 E_i}.
\end{equation}
%
Here, $\mathcal M_{a\bar a \rightarrow b \bar b}$ is the spin summed forward annihilation amplitude, S is a symmetry factor, and $f_i$ is the distribution function of the i'th species. When all the species involved follow Maxwell-Boltzmann statistics, this integral can be reduced to the well known~\cite{Gondolo:1990dk} form:
%
\begin{equation}
    C^{\rm ann}_{\rm F}(T_a) = \frac{T_a}{32 \pi^4} \int_{\max(m^2_a, m^2_b)}^\infty \id s~(s-4m^2_a) \sigma_{a\bar a \rightarrow b \bar b}K_2(\sqrt{s}/T_a),
\end{equation}
%
where $K_2(x)$ is the modified Bessel function of the second kind, and $s =(p_a+p_{\bar{a}})^{2}$ is the Mandelstam variable. If we were to fully account for quantum statistics in the initial and final states, the collision integral would no longer reduce to this simple, 1 dimensional integral. However, by neglecting final state statistics and assuming that particle  species $a$ is relativistic, we can write the forward collision integral as:
%
\begin{equation} \label{eq:annihilation}
    C^{\rm ann}_F(T_a) = \frac{T_a}{32 \pi^4} \int_{\max(m^2_a, m^2_b)}^\infty \id s~(s-4m^2_a) \sigma_{a\bar a \rightarrow b \bar b} G_{\zeta_a}(\sqrt{s}/T_a),
\end{equation}
%
with: 
%
\begin{equation}
    G_{\zeta}(x) = \int_0^\infty \id t \frac{t}{e^{x\sqrt{t^2+1}} -\zeta^2} \log \bigg( \frac{e^{x(\sqrt{t^2+1} + t)/2} + \zeta}{e^{x\sqrt{t^2+1}/2} + \zeta e^{xt/2}}\bigg).
\end{equation}
%
In this expression $\zeta_a$ is 1 (-1) if the particle is a fermion (boson). The forward energy transfer to the dark sector from annihilations is peaked around $T_{\rm D} \sim \med$ and dominated by particles lighter than $\med$, justifying the assumption of species $a$ being relativistic. Neglecting the final state statistics of species $b$ is an excellent approximation provided the density of species $b$ is much smaller than species $a$. This assumption is no longer true when the two sectors thermalize, however in the case of thermalization, the relic abundance of dark photons is insensitive to the precise value of the collision terms coupling the dark sector and Standard Model.

We approximate the total collision term by taking the backwards collision term to be the forwards collision term at $T_b$:
%
\begin{equation}
    C^{\rm ann}(T_a, T_b) = C^{\rm ann}_F(T_a) - C^{\rm ann}_F(T_b).
\end{equation}
%
The integral \ref{eq:annihilation} can not be analytically computed for a generic cross section, however it can be straightforwardly numerically integrated for a given choice of $m_a, m_b,$ and $T_a$.

For Standard Model fermions annihilating into a dark fermion pair $\chi \bar\chi$ of milli-charge $q_{\rm D}$ we use the following effective cross section, with the on-shell Z contribution subtracted off, as the on shell Z contributions will be accounted for through Z-decays
%
\begin{align}
    \sigma_{f \bar f\rightarrow \chi \bar\chi} &= \frac{4\pi \alpha^2 \qd^{2} N_C(f)}{s^3} \frac{\sqrt{s-4m^2_\chi}}{\sqrt{s-4m^2_f}} \nonumber \\
    \times &\bigg[\frac{4}{3}(2m^2_\chi +s)(2m^2_f+s)\bigg(q^2_f + \Theta(s-M^2_Z)(\frac{c^2_V + c^2_A}{4 \cos^4 \theta_{\rm W}} - \frac{c_V q_f}{\cos^2 \theta_W}) \bigg) \nonumber
    - \Theta(s-M^2_Z) ~\frac{c_V^2 + 3c^2_A}{2 \cos^4 \theta_{\rm W}}~m^2_f(s + 2m^2_\chi) \bigg]
\end{align}
%
Where $c_V$ and $c_A$ are the vector and axial couplings of the fermion species f, and $N_C$ its color factor. 
%

\subsection{Coulomb Scattering}
The collision term for an elastic scattering process is generically given by
%
\begin{align}
    C^{\rm scatt} = 2 &\int \id P \, \, S \, |\mathcal M_{a b \rightarrow a  b}|^2(E_1 - E_3) \times f_a f_b(1\pm f_a)(1\pm f_b).
\end{align}
%
This integral can be simplified down to two dimensions by defining $q = p_1 + p_3$ and $q' = p_2 + p_4$, using delta-function identities, and expanding the t-channel matrix element (assuming a generic mediator $\phi$) as:
%
\begin{equation}
    \int \id \Omega ~|\mathcal M|^2 = \frac{1}{(t-m^2_\phi)^2}\sum_{nm\lambda} c_{nm\lambda}~ (q^0)^n (q'^0)^m \frac{t^\lambda}{|\vec p|^{n+m}},
\end{equation}
%
thus yielding
%
\begin{align} \label{eq:scattering}
C^{\rm scatt}=\frac{32\pi^2}{2^7 (2\pi)^8} \,4 T_aT_b&\int_0^{\infty} \id p^0 p^0\nonumber 
 \big[\Lambda(p^0, T_a, T_b) - \Lambda(p^0, T_b, T_a)\big] \nonumber \\
&\times \sum_{nm\lambda}{c_{nm\lambda}} \frac{t^\lambda}{(t - m^2_\phi)^2}\bigg(\frac{2 T_a}{|\vec{p}|} \bigg)^n \bigg(\frac{2 T_a}{|\vec{p}|} \bigg)^m L_{n,\zeta_a}\bigg(\frac{|\vec{p}|\beta_a}{2T_a}, \frac{p^0}{2T_a} \bigg ) L_{m,\zeta_b}\bigg(\frac{|\vec{p}|\beta_b}{2T_b}, \frac{p^0}{2T_b} \bigg ).
\end{align}
%
We have have written 
%
\begin{equation*}
    \Lambda(p^0, T_a, T_b) = \frac{1}{(e^{p^0/T_a}-1)(1-e^{-p^0/T_b})},
\end{equation*} 
%
to represent the part coming from the distribution functions, and also define
\newcommand{\Li}{\rm{Li}}
%
\begin{align}
    L_{n,\zeta_a}(a, b) &= \sum_{r=0}^n \frac{n!}{(n-r)!}a^{n-r} \nonumber\big[-\zeta \Li_{r+1}(-\zeta e^{-a+b})  + \zeta \Li_{r+1}(-\zeta e^{-a-b})\big].
\end{align}
%
The integral in Eq.~\eqref{eq:scattering} cannot be computed analytically. For each dark fermion mass in our parameter space, we evaluate it on a 2-D grid of temperatures using the \texttt{vegas}~\cite{Lepage:2020tgj} Monte-Carlo integration package. Because the scattering cross section is proportional to $q_{\rm D}^2$, for each mass we evaluate the numerical integrals with $\qd = 1$ and simply rescale the integral for a particular choice of $\qd$. The spin summed matrix elements for dark fermions Coulomb scattering with Standard Model fermions is
%
\begin{equation}
    |\mathcal M_{f \chi \rightarrow f \chi}|^2 = \frac{8 \qd^2 N_c(f) Q^2_f e^4}{\big(t-m^2_\gamma(T_\gamma)\big)^2}(2(s-m^2_f-m_\chi^2)^2 + 2st + t^2).
\end{equation}
%
Above the electroweak temperature we replace $Q^2_f\rightarrow\frac{Q^2_Y}{2 \cos^4\theta_W}$ and $m_\gamma(T_\gamma) \rightarrow  m_B(T_B)$, and also include scattering with the Higgs doublet:
%
\begin{equation}
    |\mathcal M_{H \chi \rightarrow H \chi}|^2 = \frac{4 \qd^{2} e^4}{ \cos^4\theta_W\big(t-m^2_B(T_B)\big)^2}(s^2 + st - m_\chi^2(2s + t)+m^4_\chi).
\end{equation}
%

\subsection{Z-boson Decays}
The forward collision term for the decay of a massive species $a \rightarrow b \bar b$ is
%
\begin{align}
    C^{\rm decay}_{\rm F} = &\int \frac{\id^3p}{(2\pi)^3 2 E_p}\, \frac{\id^3p_1}{(2\pi)^3 2 E_1}\,  \frac{\id^3p_2}{(2\pi)^3 2 E_2} \,\,S|\mathcal M_{a\rightarrow b \bar b}|^2 (2\pi)^4 \delta^4(p - p_1 -p_2) \times f_a(p)(1 \pm f_b(p_1))(1 \pm f_b(p_2)).
\end{align}
%
Using the definition of the rest frame decay width $a\rightarrow b \bar b$ and ignoring final state statistics, this can be explicitly computed as
%
\begin{equation}
    C^{\rm decay}_F = m_a n_a \Gamma_{ a\rightarrow b \bar b}.
\end{equation}
%
Following, \cite{Adshead:2022ovo} we evaluate the inverse decay neglecting quantum statistics to get the total collision term,
%
\begin{equation}
    C^{\rm decay} = m_a \Gamma_{ a\rightarrow b \bar b}(n_{\rm eq}(T_a) - n_{\rm eq}(T_b)),
\end{equation}
%
where $n_{\rm eq}$ is the equilibrium number density of the decaying species. The width for the Z-boson decaying into a dark fermion pair $\bar \chi \chi$ is:
%
\begin{equation}
    \Gamma_{Z\rightarrow \chi \bar \chi} = \frac{\alpha \qd^{2} \tan^2 \theta_{\rm W}}{3} M_Z \sqrt{1 - \frac{4 m_\chi^2}{M^2_Z}} \bigg(1 +\frac{2m_\chi^2}{M^2_Z}\bigg)\Theta(M_Z - 2m_\chi).
\end{equation}
%

\subsection{Plasmon Decays}
The collision term of in-medium decays of plasmons is
%
\begin{equation}
    C^{\rm plas} = \sum_{\rm pol} \int \frac{\id^3 k}{(2\pi)^3 } \bigg(\frac{1}{e^{\omega/T_\gamma}-1} - \frac{1}{e^{\omega/T_{\rm D}}-1} \bigg ) ~\omega \Gamma_{\gamma \rightarrow \bar \chi\chi},
\end{equation}
%
with the plasmon decay width
%
\begin{equation}
    \Gamma_{\gamma \rightarrow \chi \bar \chi} = \frac{\alpha \qd^{2}}{3 \omega} (m^2_\gamma(T_\gamma) + 2m^2_\chi) \sqrt{1 - \frac{4m^2_\chi}{m^2_\gamma(T_\gamma)}}.
\end{equation}
%
This allows one to solve explicitly for the collision term, yielding
%
\begin{equation}
    C^{\rm plas} =  \frac{2 \alpha \qd^{2}}{3} (m^2_\gamma(T_\gamma) + 2m^2_\chi) \sqrt{1 - \frac{4m^2_\chi}{m^2_\gamma(T_\gamma)}}~\times\big(n_\gamma(T_\gamma) - n_\gamma(T_{\rm D})\big).
\end{equation}
%
As in the case of Coulomb scattering, above the electroweak temperature we take $Q^2_f\rightarrow\frac{Q^2_Y}{2 \cos^4\theta_W}$ and $m_\gamma(T_\gamma) \rightarrow  m_B(T_B)$ to properly take into account the hypercharge interactions.

\section{\label{sec:appB} Radiative Degrees of Freedom with Resonant Conversions}
In sec.~\ref{sec:level2.1}, we computed constraints on dark atomic sectors from the relic abundance of dark photons in the early Universe. We also allow, however, for resonant conversions to abruptly transfer energy density between the two sectors at later times. For the parameters of interest that will generate CMB spectral distortions, $N_{\rm eff}$ will change between BBN and recombination, and it is reasonable to ask whether this late time shift can be used to further constrain these dark photons. 

It is straightforward to show that this will not be a large effect. Imagine only one conversion, taking place at $z_{\rm res}$, whose direction of net energy transfer is into the dark sector. In this case, the CMB and dark sector energy densities some time $\delta z$ after conversion are modified
%
\begin{align}
    \rho_{\rm CMB}(z_{\rm res} - \delta z) &= \rho_{\rm CMB}(z_{\rm res}+ \delta z) - \Delta \rho_{\rm inj}, \\
    \rho_{\rm D}(z_{\rm res} - \delta z) &= \rho_{\rm D}(z_{\rm res}+ \delta z) + \Delta \rho_{\rm inj}
\end{align}
%
Now, inserting this into Eq.~\eqref{eq:Neff}, we can expand to linear order in $\Delta\rho_{\rm inj} \ll \rho_{\rm CMB},\ \rho_D+\rho_\nu$ to find
%
\begin{align}
    N_{\rm eff}(z_{\rm res}-\delta z) \simeq N_{\rm eff}(z_{\rm res}+ \delta z) \left(1 + \frac{\Delta \rho_{\rm inj}}{\rho_{\rm CMB}} + \frac{\Delta \rho_{\rm inj}}{\rho_{\rm D} + \rho_{\nu}} \right).
\end{align}
%
COBE/FIRAS constrains $\Delta \rho/\rho \lesssim 3\times 10^{-5}$, which implies that the maximum enhancement to $N_{\rm eff}$ after a conversion will contribute at a similar level. At the moment, $\Delta N_{\rm eff} \simeq \mathcal{O}(10^{-1})$, thus such a small correction will hardly alter the $N_{\rm eff}$ constraints on dark photon abundances. Large distortions ($\Delta \rho/\rho_{\rm CMB} \simeq 0.1$) are in principle possible at high ($z \geq 10^6$) redshifts, though a proper calculation of the distortion signature is beyond the aim of this work. For secluded dark photon sectors this has been considered previously in \cite{Chluba2020, Acharya2021, Chluba2024}.

\section{\label{sec:appC} Resonant Conversions for Massless Dark Photons}
As noted in Sec.~\ref{sec:level3}, a recent work by Hook \textit{et al.} \cite{Hook2025} argued that resonant conversion of \emph{massive} dark photons into SM photons in the early Universe can be rapidly curtailed by nonlinear plasma dynamics. In this appendix we summarize their arguments and further argue that they do not apply to the case of massless dark photons we consider.
In brief, their argument is as follows: the driving of longitudinal (Langmuir) oscillations induces coherent electron bulk motion and a ponderomotive force that feeds back on the local plasma density and affects the photon dispersion relation, moving the plasma masses off resonance. This allows significantly less energy transfer between the plasmas than the typical Landau-Zener analysis would suggest, weakening existing dark photon dark matter constraints derived from this mechanism.

Viable dark photon dark matter scenarios typically invoke non-thermal production (e.g., misalignment)\cite{Nelson_2011}, since thermal production is heavily constrained\cite{Fradette_2014, Ibe_2020, xu2025darkphotonsearlyuniverse}. Misalignment production yields a nearly homogeneous classical Proca field $A'_{\mu}(t, \textbf{x}) \approx A'_0\cos ({m_{A'}}t)$ with very small spatial gradients: $k\sim m_{A'}v \ll m_{A'}.$ In this regime the dark photon background is well approximated in the plasma rest frame by an almost spatially uniform oscillating electric field, 
\begin{align}
    E'&=-\dot A'-\nabla A'^0 \approx m_{A'}A'_0 \sin(m_{A'}t) + \mathcal{O}(v),\\
    B'&= \nabla \times A' \sim kA'\sim v \,m_{A'}A'\ll E'.
\end{align}
Through kinetic mixing, this coherent DPDM background sources an oscillatory SM electric field that can resonantly drive longitudinal electron-fluid oscillations. Hook \textit{et al.} argue that the resulting Langmuir wave drives the local motion of electrons relative to the ions and thus gives an oscillatory current, which they estimate parametrically as $j\sim \omega_{\rm p} E$, where $E$ is the induced SM electric-field amplitude and $\omega_{\rm p}$ is the plasma frequency. Collisional and nonlinear effects then redistribute the driven mode power to other wavenumbers and, ultimately, into electron thermal motion, leading to a modification of the local electron density. In the linear regime this back-reaction is small, but it becomes important once the driven electric field is strong enough that the work done over a plasma period is comparable to the electron thermal momentum, i.e. when
\begin{equation}
    eE\approx m_{\rm e} \omega_{\rm p} v_{\rm th}^{\rm e},
\end{equation}
where $v_{\rm th}^{\rm e}$ is the thermal speed of the electrons. The total number density of charged particles $n_{\rm tot}=n_{\rm e}+n_{\rm p}$ is proportional to $\mathrm{exp}(-e^2E^2/4m_{\rm e}^2\omega_{\rm p}^2{v_{\rm th}^{\rm e}}^2)$. Thus when the electric field energy grows larger than the thermal energy, the Landau-Zener formalism breaks down. As discussed in the main text, the resonant conversion only happens when the effective photon mass matches the effective DP mass. In the massive dark photon case, there is no dark plasma, and thus $m_{\gamma_{\rm D}}=m_{A'}$ which is simply the vacuum mass of the dark photon, a model dependent constant. The SM effective photon mass is the same as in Eq.~\ref{eq:plasma_mass} which depends on the local number density. Thus as the number density changes drastically, the SM effective photon mass will move away from the dark photon mass thus halting the resonant conversion. 

In the atomic dark sector case considered in this manuscript, however, the dark photon is massless, and thus the effective dark photon mass purely comes from plasma effects, just like the SM photon, as shown in Eq.~\ref{eq:dark_plasma_mass}. The dark photon is produced thermally, in contrast to the massive dark photon, which is produced coherently (in the $k \simeq 0$ mode). By definition, after thermalization this dark CMB will span a wide range of $k$-modes. In our case, when the resonant conversion happens, the energy transfer between the two sectors is not restricted to occurring at the plasma frequency, but instead across the whole spectrum. 

Conversions in a narrow window around the plasma frequency drive the growth of the zero mode Langmuir wave. Hook \textit{et al.} found that the largest possible fractional energy deposition before the plasma instability shuts off a resonance is roughly $\Delta\rho/\rho\sim 10^{-8}$. In order to check the validity of our analysis we can compute the fractional deposition of energy within $\delta \omega$ of the plasma frequency,
\begin{equation}
    \frac{\Delta \rho}{\rho} \simeq   \int \id z \frac{1}{\rho_{\rm CMB}} \frac{\id \rho_{\rm inj,\omega_{\rm p}}}{\id z}\mathcal{J_{\rm eff}}(z),
\end{equation}
where 
\begin{equation}
\frac{\id \rho_{\rm inj,\omega_{\rm p}}}{\id z}=   \frac{T^4_{\gamma}(z_{\rm res})}{\pi^2} \int_{\omega_{\rm p}}^{\omega_{\rm p} + \delta\omega} \id x \,x^3 \Delta n_{\gamma}(x)  \delta (z - z_{\rm res}).
\end{equation}
The largest possible spectral distortions produced in this work come at the largest milli-charge considered, e.g. $q_{\rm D} = 10^{-4}$. Even for this large dark charge, and taking an unreasonably large choice of $\delta \omega = 100 \omega_{\rm p}$, we find the fractional energy deposited into this Langmuir mode is on the order of $|\Delta \rho/\rho| \simeq 10^{-12}$. Such values are safely below the energy injection necessary to generate significant plasma instabilities in the zero mode. 

As a final comment, in all cases considered here, $T_{\rm D} < T_{\rm CMB}$. This ensures that when a resonance is passed, the CMB (and thus the standard model plasma) leaks more energy to the dark sector than it gains. Therefore, the standard model plasma should never experience instabilities due to energy injection, as no injection actually occurs. Of course, in these scenarios the dark plasma may develop instabilities, but as we have shown above, the fractional injection near the plasma frequency is negligible and the usual formalism should remain valid.

\bibliographystyle{apsrev4-1}
\bibliography{Lit}

\end{document}